\documentclass[epj]{svjour}
\usepackage{graphics}
\begin{document}
\title{Exploration of Scale-Free Networks}
\subtitle{Do we measure the real exponents?}
\author{Thomas Petermann \and Paolo De Los Rios
}                     % Do not remove
\institute{
Laboratoire de Biophysique Statistique, ITP-FSB, 
Ecole Polytechnique F\'ed\'erale
de Lausanne, 1015 Lausanne, Switzerland.}
\date{Received: date / Revised version: date}
\abstract{
The increased availability of data on real networks has favoured an 
explosion of activity in the elaboration of models able to reproduce both
qualitatively and quantitatively the measured properties.
What has been less explored is the reliability of the data, and whether
the measurement technique biases them. Here we show that tree-like explorations
(similar in principle to {\it traceroute}) can indeed change the measured 
exponents of a scale-free network. 
 % end of PACS codes
} %end of abstract
\authorrunning{T. Petermann and P. De Los Rios}
\titlerunning{Exploration of Scale-Free Networks}
\maketitle
\section{Introduction}
\label{intro}
In recent years networks have become one of the most promising
frameworks to describe systems as diverse as the Internet and the WWW, email
and social communities, distribution systems,
food-webs, protein interaction, genetic and metabolic networks~\cite{AB02}.
The collected data have allowed the discovery of many important properties: 
in particular two of them have become prominent, 
namely the {\it small-world}~\cite{WS98} and {\it scale-free} features~\cite{BA99}. 
{\it Small-world} implies that 
the average distance between nodes of the network increases at most 
logarithmically with the number of nodes, and formalizes the
concept of "six degrees of separation" typical in social contexts.
{\it Scale-free} refers to the
lack of an intrinsic scale in some of the properties of the network.
In particular, the quantitiy that has been most thoroughly studied
is the degree (or connectivity) distribution: the degree $k$ of a node is the 
number of other nodes it has links to (here we do not distinguish between
directed and undirected links), and the degree distribution $P(k)$ is simply 
the histogram of the number of nodes with a given degree $k$.
Scale-free networks exhibit a power-law behavior of the distribution 
$P(k) \sim k^{-\gamma}$, with $\gamma$ values often between $2$ and $3$~\cite{AB02}.
The small-world and scale-free  properties turned out being quite ubiquitous
and some general, qualitative, models to explain their appearence have 
been put forward. At the same time various versions of these models have 
been also proposed in order to capture also the detailed values of some 
quantities, such as the exponent $\gamma$. Yet, as this new field is slowly
coming of age, and as a consequence it is also becomeing
more quantitative, an analysis of the data, and of their reliability, 
is due. The main problem that should be addressed is whether the data we are
using have been skewed somehow by the detection method. 
In the lack of such analysis on real data and methods,
we propose to work on syntetic models and data to explore their robustness
in some simple test case. In the next section 
we address the tree-like exploration technique and discuss how it can bias the measurements,
and in the third section we show that a random graph can be distorted by the exploration
so to look like a SF one: in this case the exponent $\gamma$ is completely spurious.
\section{Tree-like Exploration of Scale-Free Networks}
\label{sec:1}
Scale-free (SF) networks can be explored in many different ways. One of the most popular
methods, that has been extensively used for example for the Internet, is a 
sort of tree-like exploration implemented by the recursive use of the {\it traceroute} command.
In short, {\it traceroute} finds a path (usually a short one, but not necessarily the shortest)
from the node where the command is executed to another given node. By repeating the procedure
asking {\it traceroute} to find paths to all other possible nodes (addressed by their 
IP number), one ends up with a representation of the Internet that shows just a small
amount of loops. This is due to the fact that {\it traceroute} mostly uses the same
paths: if a node $D$ can be reached from $A$ through both $B$ and $C$, {\it traceroute}
most of the times detects only one of them. Actually, chances are that {\it traceroute} 
can find more than a single path if traffic over an already discovered one is so high
that it becomes more convenient to switch to a different path. 
Data collected with this technique have shown that degrees in the Internet are
distributed according to a power-law with exponent $\gamma \simeq 2.2\pm 0.1$
~\cite{MCP}.
In order to analyze the effects of a tree-building exploration algorithm on 
SF networks, we have synthesized our own networks according to two different
models: the Barabasi-Albert (BA) model~\cite{BA99}, and the hidden variable model
~\cite{CCDM02}.
 
The BA model describes the growth of a network as new nodes are added at a constant
rate, and they connect to older nodes in the network according to the
preferential attachment rule. Preferential attachment means that an old node has a 
probability proportional to its degree of aquiring a connection from a new one.
It is useful to recall a simple derivation of the degree distribution 
starting from these two simple rules, growth and preferential attachment.
The rate of change of the degree $k_i$ of node $i$ is
\begin{equation}
\frac{dk_i(t)}{dt} = m \frac{k_i(t)}{2mt}
\label{basic rate equation}
\end{equation}
where $m$ is the number of connections that a new node establishes with older ones,
and the denominator in the {\it right hand side} of Eq.\ref{basic rate equation}
represents the sum over all the degrees of the network. Eq.\ref{basic rate equation}
has the simple solution $k_i(t) = m(t/\tau_i)^{1/2}$, where $\tau_i$ is the time at which
node $i$ entered the network. Since the relation between $k_i$ and $\tau_i$ is monotonous
we can classify nodes according either to their degree $k$ or to their age $\tau$.
As a consequence we can apply the usual formula to transform probability distributions:
$P(k)dk = \rho(\tau) d\tau$. Since nodes enter the network at a constant rate, 
we have $\rho(\tau) = const$ and therefore $P(k) \sim k^{-3}$.
As mentionent above for the Internet, the exponent $\gamma$ is in general
not equal to the BA prediction $\gamma=3$; yet it has been shown that, as long
the attachment rate in (\ref{basic rate equation}) is asymptotically linear in $k$,
the distribution $P(k)\sim k^{-\gamma}$ with $2<\gamma<\infty$ depending on the
pre-asymptotic behavior~\cite{KRL}.
Another important feature of the BA model, that we are going to exploit for our
analytical approach, is the lack of correlations between the degree of a node and the 
degrees of its neighbors. This is best represented through the average neighbor degree
$k_{nn}(k)$, that is the average degree of the neighbors of a node of degree $k$:
this quantitiy is essentially constant for the BA model.

\begin{figure}
\resizebox{0.6\textwidth}{!}{\includegraphics{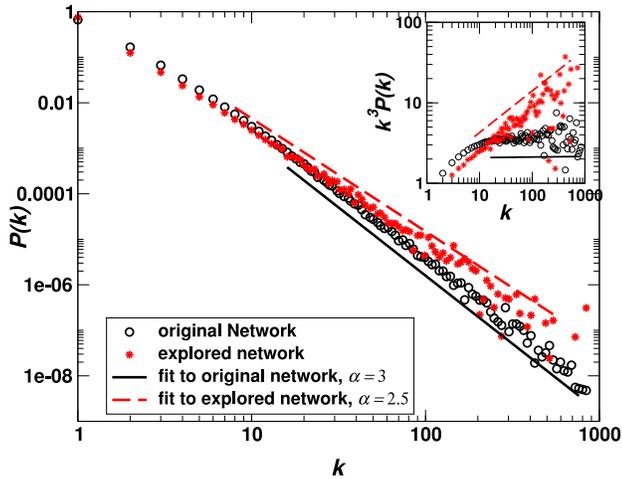}}
\caption{Degree distribution for a Barab\'asi-Albert network grown with
$m=1$, with $10^5$ nodes. Circles: original network; stars: explored network
with $p=0.5$. The best fit to the original network is with $\gamma \simeq 3$,
and to the explored network with $\gamma \simeq 2.5$. Inset: rescaled degree distribution
$k^3 P(k)$, such that the data for the original network are constant, and the
residual power-law behavior of the explored network is more evident.}
\label{Fig1}
\end{figure}

The first step in our analysis is to build a BA network, whose degree distribution
is shown in Fig.\ref{Fig1}. Then, starting from a node (we choose a highly connected
node) we begin our exploration procedure: 1) each edge connecting that node to its
neighbors is followed with probability $p$; edges that are lost at this stage are lost
forever 2) from each of the reached nodes, repeat step 1) until no new nodes are reachable.
In this procedure, edges to nodes that have already been reached are not
followed. The result of this algorithm is a network that has fewer nodes than the original one, 
and, on the average, fewer links per node, and that is, topologically, a tree. 
The intuitive result would be that 
every node sees just a fraction $p$ of its edges, so that all degrees should be reduced of 
a factor $p$, without consequences on the power-law behavior of $P(k)$.
Actually the effects of the probability $p$ are much more dramatic. As it can be seen 
from our simulations (Fig.\ref{Fig1}), the measured exponent $\gamma_m$ actually changes.
For a network grown with $m=1$ and explored with $p=0.5$ the measured exponent is close 
to $\gamma_{m} = 2.5$. We can 
therefore wonder whether this is a crossover effect, and the correct exponent is recovered
for very large networks, or whether this change of exponent is real.
A simple analytical argument in favour of this second interpretation can be formulated 
using the lack of correlations in the BA model. Indeed, since there is no correlation
between the degree of a site and the degrees of its neighbors, due to (\ref{basic rate 
equation}) there is no correlation between the age of a node and the age of its neighbors.
This allows us to look at exploration during the growth of the network. In particular
we can say that, in a growing network formalism, any time a new node is added to the network,
we label it as reachable if it connects to at least a reachable site through a followed
connection (with probability $p$). We assume that the first site is reachable. 
Then, the density of reachable nodes at time $t$ is given by
\begin{equation}
\frac{dN(t)}{dt} = 1-\left(1-p\int_0^t \frac{dN(t')}{dt'} q(t') dt'\right)^m
\label{rate equation for Number}
\end{equation}
where $q(t')$ is the probability to choose a node introduced in the network
between $t'$ and $t'+dt'$: the 
preferential attachment rule translates to $q(t') = 1/[2(t\cdot t')^{1/2}]$
(this trick is similar to assigning to each node a hidden variable 
corresponding to the time $t'$ at which it entered the network, 
with a connection probability that depends on the hidden variables
of both the new and old nodes; for more details see below~\cite{BPS03,CBS03}).
Since $N(t)$ can grow at most linearly, we make the assumption that $dN(t)/dt \sim t^{\alpha}$
with $\alpha$ expected to be negative.
After some algebra, and keeping only the leading terms, we find $\alpha = (mp-1)/2$:
as long as $mp<1$ the density of reachable nodes decreases in time.
Then, the measured degree distribution can be again obtained from the relation
$P_m(k)dk = \rho(\tau) d\tau$, with $\rho(\tau) \sim \tau^{\alpha}$, from which we obtain
$P_m(k) \sim k^{-\gamma_m}$, with $\gamma_m = 2+mp$
For $m=1$, $p=0.5$ we have $\gamma_m=2.5$, in agreement with simulations.
We expect therefore that, as long as $mp<1$, the measured exponent
could be different from the real one.

\begin{figure}
\resizebox{0.6\textwidth}{!}{\includegraphics{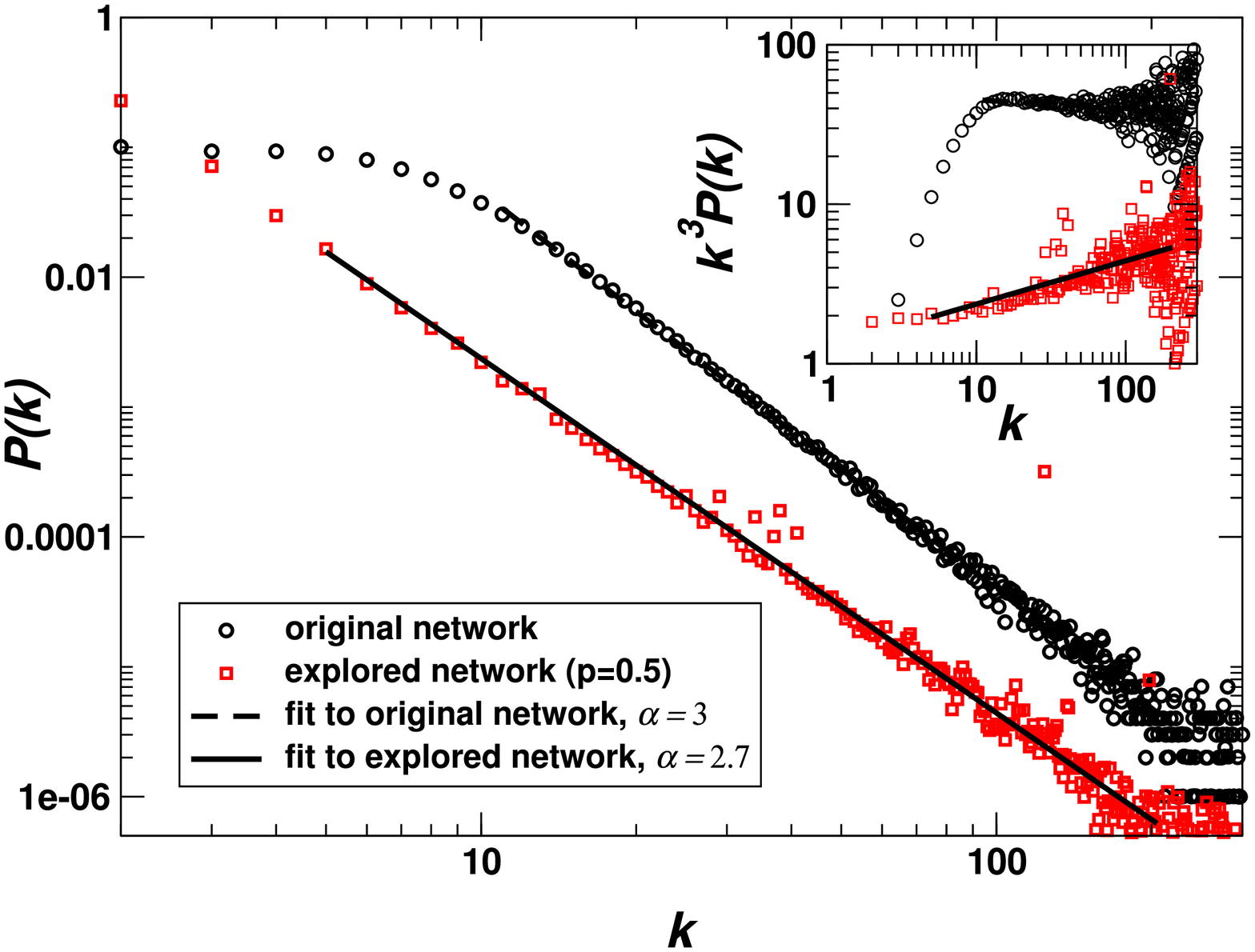}}
\caption{Degree distribution for a hidden variables network with
$p(x) \sim x^{-3}$ and $q(x,x') \propto x \cdot x'$; $10^5$ nodes.
Circles: original network; squares: explored network
with $p=0.5$. The best fit to the original network is with $\gamma \simeq 3$,
and to the explored network with $\gamma \simeq 2.7$. Inset: rescaled degree distribution
$k^3 P(k)$, such that the data for the original network are constant, and the
residual power-law behavior of the explored network is more evident.}
\label{Fig2}
\end{figure}

To check whether the distortion of the exponent is a feature only of BA networks, we
have also studied networks generated according to the hidden variables model.
Hidden variable networks are characterized by a quantity $x$ (the "fitness") assigned to 
every node and taken from some probability distribution $p(x)$; every pair of nodes $i$ and
$j$ is connected then with a probability $q(x_i,x_j)$. As a consequence the average
degree of a node of fitness $x$ is
\begin{equation}
k(x) = N \int_U q(x,x') p(x') dx'
\label{average degree}
\end{equation}
where $N$ is the number of nodes in the network and $U$ is the support of $p(x)$. 
Eq.\ref{average degree} gives a relation 
between $k$ and $x$ that can be used in $P(k)dk = p(x)dx$ to obtain $P(k)$.
Suitable choices of $p(x)$ and of $q(x,x')$ give SF networks. Here we
use the same examples provided in~\cite{CCDM02}, namely Zipf and exponential
fitness distributions. Zipf distributed fitnesses are inspired by the idea that 
many quantities such as personal wealth, company size, city population and others 
are power-law distributed~\cite{Zipf}. In this case a connection probability 
$q(x,x') \sim x \cdot x'$ ensures that the resulting network is SF. 
Using $p(x) \sim x^{-3}$ we obtain $P(k) \sim k^{-3}$ and 
in Fig.\ref{Fig2} we show that also this network's exponent has changed
following the exploration, with a value $\gamma_m \simeq 2.7$.

\begin{figure}                          
\resizebox{0.6\textwidth}{!}{\includegraphics{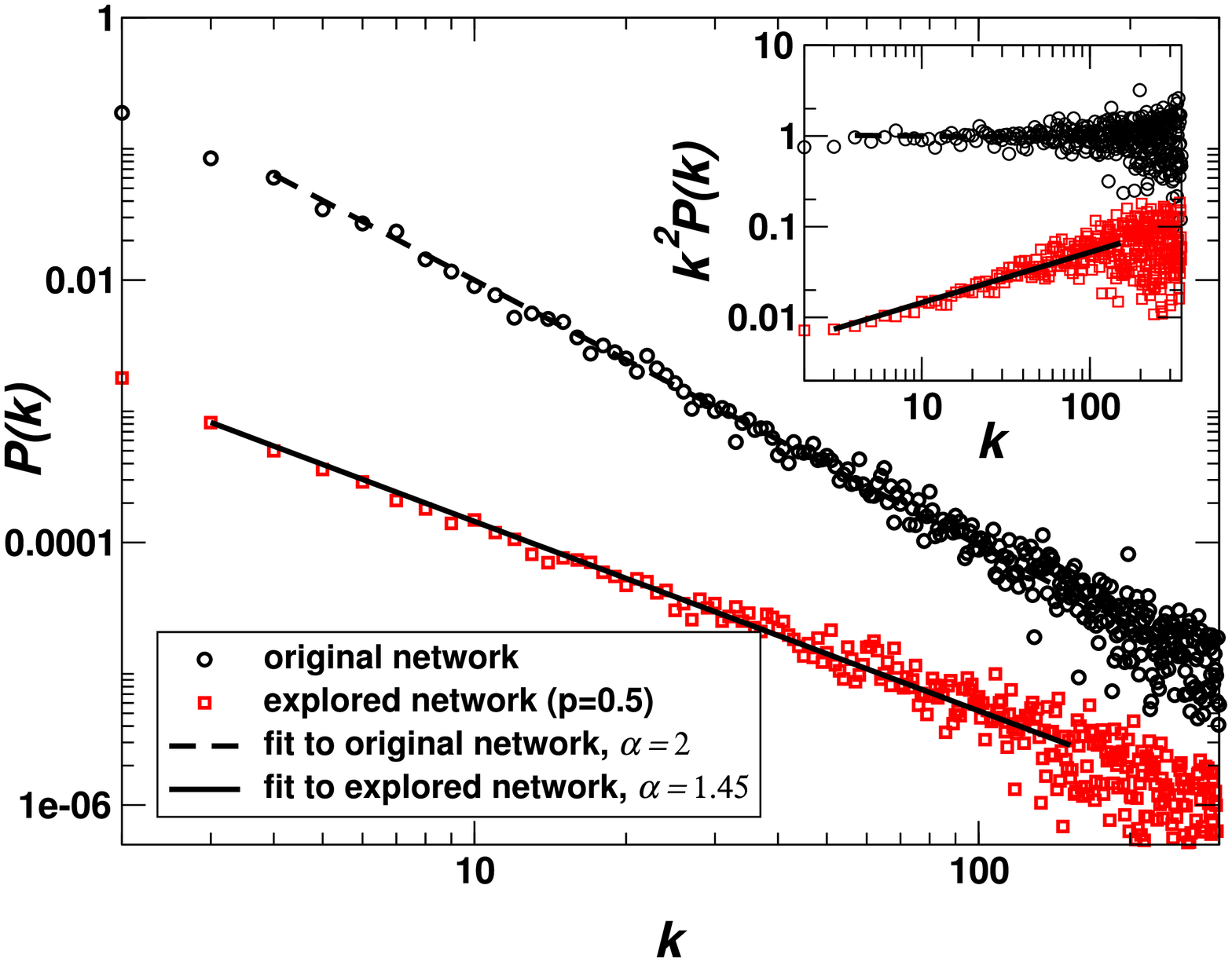}}
\caption{Degree distribution for a hidden variables network with
$p(x) \sim e^{-x}$ and $q(x,x') = \theta(x+x'-x_c)$; $10^5$ nodes.
Circles: original network; squares: explored network
with $p=0.5$. The best fit to the original network is with $\gamma \simeq 2$,
and to the explored network with $\gamma \simeq 1.45$. Inset: rescaled degree distribution
$k^2 P(k)$, such that the data for the original network are constant, and the
residual power-law behavior of the explored network is more evident.}
\label{Fig3}
\end{figure}

Next we look at SF networks obtaned using an exponential fitness distribution
$p(x)=exp(-x)$ and a connection probability $q(x,x') = \theta(x+x'-x_c)$,
that is, a link between two nodes of fitness $x$ and $x'$ is present only if the
sum $x+x'>x_c$. Eq.\ref{average degree} yelds $k(x) = N e^{-x_c} exp(x)$ and
as a consequence $P(k) \sim k^{-2}$ (see Fig.\ref{Fig3})~\cite{CCDM02}. 
The tree-like exploration
of this network shows that, also in this case, the measured exponent can change,
$\gamma_m \simeq 1.5$.
We do not have at the moment an analytical derivation of the measured exponents. Indeed,
degree-degree correlations between nearest neighbors and the lack of an
explicit time evolution hinder the formulation of some equations similar to
(\ref{rate equation for Number}).

Interestingly, in all cases we have analysed, the measured exponent $\gamma_m < \gamma$,
an indication that the exploration process penalizes nodes with small degree
with respect to nodes with large degree. This is reasonable, since a node with
few connections has fewer paths reaching it (and some bottlenecks, since all these
paths have ultimately to flow through its few connections) than a high degree node.
The final result is therefore that high degree nodes are fairly well represented in 
the final distribution, whereas the number of nodes with few connections is 
underestimated. This intuitive picture rationalizes our numerical and analytical
finding that the measured exponent is smaller than the real one.
 
\section{Edge-Picking Exploration of Erd\"os-R\'enyi and Complete Graphs}
\begin{figure}
\resizebox{0.55\textwidth}{!}{\includegraphics{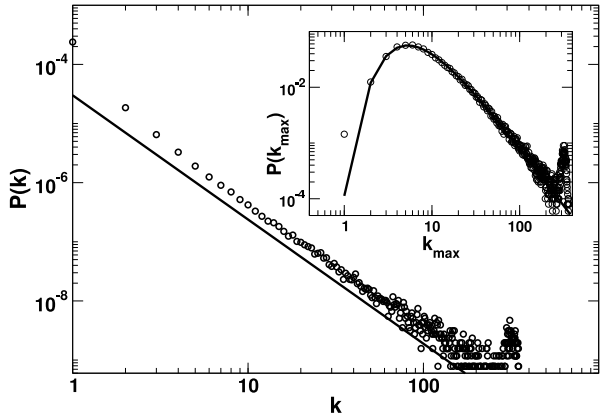}}
\caption{Main panel: Degree distribution for ER networks ($p=0.025$) of $N=12800$ nodes,
where every node has been assigned a variable $x$ taken from an exponential distribution.
A link between $i$ and $j$ is detected only if 
$x_i+x_j>x_c=13$; data are averaged over $10000$ realizations.
The solid line is $k^{-2}$.
Inset: maximum degree distribution for the $10000$ networks; the solid line is a Frechet
distribution of exponent $2.1(1)$.}
\label{Fig4}
\end{figure}

The last example is also a good example of how other link
detection techniques can change the apparent topological properties of
networks in an even more dramatic way, in particular if the probability to detect
a link depends on some intrinsic properties of the nodes it connects.
Indeed, we can interpret the generation of SF networks from hidden variables
as the result of the exploration of complete or Erd\"os-R\'enyi~\cite{ER} (ER) networks.
Starting from a complete or ER graph, we assign to each node $i$ a variable $x_i$ taken
from a probability distribution $p(x) = exp(-x)$. Then, we prune the graph
by discarding (that is, they are not detected)
all those edges that join nodes $i$ and $j$ such that $x_i+x_j < x_c$,
with $x_c$ a properly chosen threshold. This is tantamount to say that, during
the exploration of the graph, the edge between nodes
$i$ and $j$ is detected with probability $q(x_i,x_j) = \theta(x_i+x_j-x_c)$,
which brings us back to the formalism used above, that shows that the resulting
network is scale-free with degree distribution $P(k)\sim k^{-2}$ (data shown in
Fig.\ref{Fig3}, black symbols, refer to the procedure over a complete graph).
Fig.\ref{Fig4} (main panel) shows the results for a starting graph that is an Erd\"os-R\'enyi network
of $12800$ vertices with a connection probability $p=0.025$, 
corresponding to a graph with average degree $<k>=320$;
the threshold $x_c=13$. The resulting degrees are power-law distributed with exponent $\gamma \simeq 2$
(black thick line). The peak for large values is what is left of the Poisson
distribution of the underlying ER network: whenever a node has a variable $x>x_c$,
all of its connections are detected, and its degree is not distorted.
Since in SF networks a special role is played by hubs, that is, nodes with a very large degree,
we checked whether the hubs of the explored network fall into the Poisson cutoff or in the
power-law distributed part. The result, shown in the inset of Fig.\ref{Fig4} clearly shows that the
distribution of the maximum degree nicely obeys a Frechet distribution $P_{max}(k) = (\gamma -1)
k^{-\gamma} exp(-k^{-\gamma +1})$~\cite{MAA02}, with $\gamma = 2.1(1)$. Such a Frechet distribution
is indeed the expected distribution of the maxima of set of variables
taken from a power-law distribution $k^{-\gamma}$. This implies that, apart from a $1\%$
of the networks (we show the distribution of $k_{max}$ over $10000$ networks), the maximum degree is
almost always drawn from the power-law part of the degree distribution, an indication that typical networks
can be considered genuinely scale-free.
 
\section{Conclusions}
The increase in the amount of real network data is prompting the community to study
networks in more detail and to elaborate models able to predict qualitatively,
but also quantitatively, the measured properties. Yet, before taking these data by face value,
a thorough investigation of the measurement techniques is necessary to ascertain if
and what kind of data distortion they could introduce. This is customary in physics,
where systematic errors have always to be taken into account and possibly to
be corrected, and the same kind of attention should be paid also to data
from different disciplines.
In this work we have shown, through simple examples, that tree-like explorations,
inspired by the {\it traceroute} command, can indeed skew the data so that the measured
exponent of the degree distribution of a scale-free network can change with respect to the
real one. In the simplest case (BA networks), simulations and simple analytical
arguments agree with each other.
We have also shown that a recently proposed model of SF networks based on hidden variables can
be interpreted as an exploration technique leading to the appearence of power-law degree distribution
where the underlying network is topologically much simpler.
\\
 
This work has been supported by
the FET Open Project IST-2001-33555 COSIN,
and by the OFES-Bern (CH).

\end{document}